\documentclass[pra, amsmath, reprint, amssymb, aps, superscriptaddress, twocolumn]{revtex4-1}

\usepackage[utf8]{inputenc}
\usepackage{graphicx}
\usepackage{xcolor}
\usepackage{amsmath}
\usepackage{amsfonts}
\usepackage{fixmath}
\usepackage{physics}
\usepackage{graphicx}
\graphicspath{ {fig/} }
\usepackage{dsfont}
\usepackage{braket}
\usepackage{svg}
\usepackage{float}
\usepackage{caption}

\usepackage{multirow}
\usepackage{booktabs}

\usepackage{upgreek}

\usepackage[colorlinks,linkcolor=blue,anchorcolor=red,citecolor=green]{hyperref}

\allowdisplaybreaks

\begin{document}


\title{Equivalence and Divergence of Imaginary-Time Evolution and Gradient Descent for Gaussian Variational States}

\author{Yash Palan}
\email[]{palan@itp.uni-frankfurt.de}
\affiliation{ Institut f\"ur Theoretische Physik, Goethe-Universit\"at, 60438 Frankfurt a.M., Germany}

\date{\today}

\begin{abstract}
    Imaginary-time evolution (ITE) is one of the most widely used numerical techniques for obtaining ground states of many-body Hamiltonians. In this work, we compare ITE with gradient descent (GD) within the framework of Gaussian wavefunction ans{\"a}tze. We show that while ITE and GD are formally equivalent for fermionic systems, GD exhibits consistently faster convergence for bosonic systems, challenging the common assumption of their complete equivalence.
\end{abstract}


\maketitle
\section{Introduction}
Extracting the ground state of many-body Hamiltonians is a central problem in condensed matter physics. Analytical approaches often become intractable in the presence of strong correlations, motivating the development of a range of numerical methods. A few among these are the density matrix renormalization group (DMRG) \cite{DMRG_paper_1,DMRG_paper_2,DMRG_paper_3}, quantum Monte Carlo (QMC) \cite{Costa:QMC_phonon_dispersion,QMC_paper_2,QMC_paper_3}, dynamical mean-field theory (DMFT) \cite{DMFT_paper_1,DMFT_paper_2}, density functional theory (DFT) \cite{DFT_paper_1}, and imaginary-time evolution (ITE) \cite{Demler:Kondo_model,Shi2018:Variational_study_of_fermionic_and_bosonic_systems,Wang:Hubbard_Holstein_model}.  \\
In parallel to ITE, gradient descent (GD)—the foundational algorithm for modern machine learning—has proven to be a highly effective general-purpose optimization technique. Since both ITE and GD can be formulated as iterative minimization processes, it is natural to inquire about their precise relationship and comparative performance in variational quantum simulations.  \\
In this work, we systematically analyze the connection between ITE and GD, demonstrating that although they yield equivalent dynamics for single-particle systems, their behavior diverges in many-body cases. In particular, we find that GD converges faster than ITE for bosonic systems, challenging the conventional assumption of full equivalence between the two. \\
The remainder of the paper is organized as follows. Section 2 revisits the ITE–GD correspondence for single-particle Hamiltonians. In Section 3 we extend the analysis to systems of bosons and fermions using a Gaussian-state ansatz, as well as providing a comparison between the two for these two types of systems. 
\section{Single particle Hamiltonians}
Imaginary time is a strong mathematical trick with extreme utility in describing physics, as seen from the literature in quantum field theory. Imaginary time evolution borrows the trick of making a Wick's rotation, defined as $t\rightarrow it$, to the Schr{\"o}dinger's equation. This forces the evolution to be damp the energy eigenstates rather than induce a phase shift, effectively modeling wavefunction loss. However, normalising the wavefunction  after each iteration preserves the lowest lying eigenstate and enhances its weight with time. This selection of the lowest lying energy state is the core idea behind Imaginary time evolution (ITE). The modified Schr{\"o}dinger's equation for ITE is \cite{Shi2018:Variational_study_of_fermionic_and_bosonic_systems} 
\begin{align}\label{eq:ITE_eqaution_of_motion_sph}
    \partial_{\tau}\ket{\psi} = -\left(\hat{H} - E\right) \ket{\psi}\:\text{where}\:E= \bra{\psi}\hat{H}\ket{\psi},
\end{align}
where the subtraction of the average energy deals with the normalization condition.
The ground state in the limit $t \rightarrow \infty$. Another well established technique for determination of the ground state is the variational principle, which asserts that the quantum state that minimizes the energy expectation value is the ground state.
Gradient descent (GD) is a well known technique for minimization of convex energy functions, which naturally raises the question of whether a link exists between these two techniques. In the single particle case, the two methods are exactly the same, i.e.\ both result in the same update equations \cite{SciPostPhys:scalable_ITE_with_NN_quantum_states}. For completion's sake, we provide a brief proof of the same. The gradient of the energy functional 
\begin{align}
    E 
    &= \int \: dx \:\:\psi^{\dagger}(x)\left[-\frac{\hbar^2}{2m}\partial^2_{x} + V(x) \right] \psi(x) \\
    &= \int dx \sum^{2}_{i=1}\left[-\frac{\hbar^2}{2m}\kappa_{i}\partial^2_{x}\kappa_i + V\kappa_{i}^2 \right],
\end{align}
with respect to the quadrature functions $\kappa_1,\kappa_2:\mathds{R}^d\rightarrow \mathds{R}^d$ (defined by $\psi(x) = \kappa_1(x) + i\kappa_2(x)$) can be computed as 
\begin{align}
     \fdv{E}{\kappa_i(x)}= \left[-\frac{\hbar^2}{m} \partial^2_{x} \kappa_i + 2V\kappa_i\right], \quad i =1,2.
\end{align}
Ensuring the normalisation condition $\int dx \sum_{i}\kappa^2_{i}(x)=1$ hold true during the descent demands projecting the gradient to the appropriate tangent space. This can be done by extracting the normal vector to the tangent space 
\begin{align}
    \va{f} = \left(\fdv{\kappa_1},\fdv{\kappa_2}\right)\int dx\sum_{i}\kappa^2_{i} = 2\left(\kappa_{1}(x),\kappa_{2}(x)\right).
\end{align}
The projected gradient is then
\begin{align}
    \va{v}_{\parallel} = \left(\pdv{E}{\kappa_i},\pdv{E}{\eta_i}\right) - \lambda \hat{f} 
\end{align}
where $\lambda = \int dx\sum^2_{i=1} \kappa_{i} \pdv{E}{\kappa_{i}}    = 2 E$. This provides us with the update rules for the quadrature functions and wavefunction as 
\begin{align}
    \delta \kappa_i &= -\alpha \left(-\frac{\hbar^2}{m} (\partial^2_{x} \kappa_i) + 2V\kappa_i - 2E \kappa_i   \right) \\
    \delta \psi &= - 2\alpha\left[ -\frac{\hbar^2}{2m} \partial_{x}\psi + V \psi - E\psi  \right] \\
    & = - \alpha'\left[\hat{H}-E\right]\left[\psi(x)\right]
\end{align}
where $\alpha$ determines the step size.
This depicts the equivalence for ITE and GD for a first quantised Hamiltonain based variational technique\cite{SciPostPhys:scalable_ITE_with_NN_quantum_states}. In addition, the above also guarantees convergence GD to the global energy minima.
\section{Many body Hamiltonians: Gaussian states}
The applicability of imaginary-time evolution (ITE) extends well beyond single-particle Hamiltonians. It is extensively employed to determine ground states of interacting many-body systems, including Bose–Einstein condensates (BECs) \cite{Liang2018}, superconductors \cite{Wang:Hubbard_Holstein_model}, Kondo systems \cite{Demler:Kondo_model}, and polaron models \cite{}. A common strategy in such systems is to perform a suitable canonical transformation and restrict the resulting dynamics to a Gaussian manifold, on which the transformed Hamiltonian is minimized \cite{Shi2018:Variational_study_of_fermionic_and_bosonic_systems}.
\subsection{Gaussian state}
Consider a system with $N_b$ bosons ($b^{\dagger}_{i}$) and and $N_f$ fermions ($c^{\dagger}_{i}$). It is generally easier to define the Gaussian ansatz in the quadrature operators ($x_i,p_i$) and Majorana basis ($a_{1,i}, a_{2,i}$), which are defined as $x_{i} = b^{\dagger}_i+b_i$, $p_i = i(b^{\dagger}_i-b_i)$, and $a_{1,i} = c^{\dagger}_{i} + c_{i}$, $a_{2,i} = i(c^{\dagger}_{i} - c_{i})$, along with their associated vectors 
\begin{align}
    R &= \left(x_{1},x_2,\ldots,x_{N_b},p_1,p_2,\ldots,p_{N_b}\right)^{T} \\
    A &=  \left(a_{1,1},a_{1,2},\ldots,a_{1,N_f},a_{2,1},a_{2,2},\ldots,a_{2,N_f}\right)^T.
\end{align}
The Gaussian ansatz is simply given by as $\ket{\Psi_{GS}} = U_{GS} \ket{0}$ where \cite{Shi2018:Variational_study_of_fermionic_and_bosonic_systems}
\begin{align}
    U_{GS} &= e^{i\theta} e^{\frac{i}{2}R^{T}\sigma \Delta_{R}} e^{-\frac{i}{4}R^{T}\xi_{b}R} e^{\frac{i}{4}A^{T}\xi_{m}A}, \\
     \sigma &= \begin{bmatrix}
                0 & \mathds{1}_{N_b \times N_b} \\
                -\mathds{1}_{N_b \times N_b} & 0
        \end{bmatrix},
\end{align}
where $\ket{0}$ is the vacuum state and $\xi_b$ ($\xi_m$) is a real, symmetric (antisymmetric) matrix. Each family of Gaussian states can be uniquely defined by the Displacement vector ($\Delta_R$), the Bosonic covariance matrix ($\Gamma_b$) and the Majorana covariance matrix ($\Gamma_m$)
\begin{align}
    (\Delta_R)_{i} &= \bra{\Psi_{GS}}R_i\ket{\Psi_{GS}} \\
    (\Gamma_b)_{i,j} &= \frac{1}{2}\bra{\Psi_{GS}}\acomm{ R_{i}-(\Delta_{R})_i}{ R_{j}-(\Delta_{R})_j}\ket{\Psi_{GS}} \\
    (\Gamma_m)_{i,j} &= \frac{i}{2}\bra{\Psi_{GS}}\comm{A_{i}}{A_{j}}\ket{\Psi_{GS}}.
\end{align}
For more details we refer the reader to \cite{Shi2018:Variational_study_of_fermionic_and_bosonic_systems}.
\subsection{Imaginary time evolution equations}
Taking the above ansatz, it is possible to find the equations of motion for the displacement vector and the covariance matrices by projecting the ITE equation of motion (equation \ref{eq:ITE_eqaution_of_motion_sph}) onto the single particle bosonic subspace, 2 particle bosonic subspace and the two particle fermionic subspace. This process gives us the following equations of motion \cite{Shi2018:Variational_study_of_fermionic_and_bosonic_systems}
\begin{align}
    \partial_{\tau} \Delta_{R} &= -\Gamma_b h_{\Delta} \label{eq:ite_many_body_displacement} \\
    \partial_{\tau}\Gamma_b &= \sigma^{T} h_{b} \sigma-\Gamma_bh_b\Gamma_b \label{eq:ite_many_body_bosonic_covariance} \\
    \partial_{\tau}\Gamma_m &= -h_m - \Gamma_m h_m \Gamma_m. \label{eq:ITE_fermionic_covariance}
\end{align}
where the functional derivative matrices are
$(h_{\Delta})_i = 2\pdv{E}{(\Delta_{R})_{i}} $, $(h_{b})_{i,j} = 4\pdv{E}{(\Gamma_b)_{i,j}} $ and $    (h_{m})_i = 4\pdv{E}{(\Gamma_{m})_{i,j}} $.

\subsection{Gradient Descent}

\subsubsection{Fermions}
We begin by examining the gradient descent (GD) equations for the fermionic sector. For a pure state, the covariance matrix $\Gamma_m$ is an element of the orthogonal group $O(2N_f)$ \cite{Shi2018:Variational_study_of_fermionic_and_bosonic_systems}. Given a Hamiltonian $H$, the gradient is 
\begin{align}
    \frac{h_m}{4} = \frac{\partial E}{\partial \Gamma_m}.
\end{align}
A naive update rule is $\Gamma_m \rightarrow \Gamma_m - \alpha h_m$, which would, however, take $\Gamma_m$ outside $O(2N_f)$. To maintain the orthogonality property, we need to project the gradient onto the tangent subspace of the orthogonal group. \\
The group $O(2N_f)$ is a Lie group, and hence a differentiable manifold.  
To obtain the tangent space, we define a smooth curve
\begin{align}
    \phi(t) = \Gamma_m e^{tA}, \qquad A \in \mathfrak{o}(2N_f),
\end{align}
where $\mathfrak{o}(2N_f)$ is the associated Lie algebra of real antisymmetric matrices.  
The tangent vector at $\Gamma_m$ is
\begin{align}
    M = \left. \frac{d\phi(t)}{dt} \right|_{t=0}.
\end{align}
Using $A = \Gamma_m^{-1} M$ and $A^T = -A$, we obtain the defining condition for the tangent space:
\begin{align}
    M \Gamma_m^T + \Gamma_m M^T = 0.
\end{align}
A projected update satisfying this condition is given by
\begin{align}
    \Gamma_m \rightarrow \frac{\alpha}{2}\left[\,h_m - \Gamma_m h_m^T (\Gamma_m^{-1})^T\,\right] = \frac{\alpha}{2}M''=M',
\end{align}
with $\alpha \in \mathds{R}$ chosen so that the inner product between $h_m$ and $M'$ is positive i.e.\ $\Tr(h_m^T M') > 0$. This condition is easily fulfilled by choosing
\begin{align}
    \alpha 
    &= \mathrm{sgn}\!\left( \frac{\Tr(h_m^T M')}{\Tr(h_m^T h_m)} \right) = \text{sgn}\left(\Tr(h_m^T M')\right)
\end{align}
For a pure state, $\Gamma_m$ is orthogonal and thus diagonalizable.  
Since $\Gamma_m^2 = -\mathds{1}$, its eigenvalues satisfy $\lambda^2 = -1$, implying $\lambda = \pm i$.  
We may therefore write $\Gamma_m = U^{\dagger} D U$, where $U$ is unitary and $D = \mathrm{diag}(\pm i)$.  
Defining $A = U h_m U^{\dagger}$ and $D' = iD \in \mathds{R}$, we find
\begin{align}
    \Tr(h_m^{T} M')&=\Tr(h_m^{\dagger} M')\\ 
    &= \tfrac{1}{2}\!\left[\Tr(A^{\dagger} A) + \Tr(A^{\dagger} D' A D')\right] \nonumber\\
    &= \tfrac{1}{2} \sum_{i,j} |A_{ji}|^2 (1 - d'_j d'_i) \ge 0,
\end{align}
since $\mathrm{diag}(D') = (d'_1, d'_2, \ldots, d'_{2N_f})$ with $d'_i = \pm 1$, and $h_m$ is a real matrix implying that $h^{\dagger}_m=h^{T}_m$. The above means that we can choose $\alpha = 1$. The gradient descent update rule for $\Gamma_m$ becomes
\begin{align}
    \Gamma_m &\rightarrow \Gamma_m - \frac{\kappa}{8}\!\left[\,h_m - \Gamma_m h_m^T (\Gamma_m^{-1})^T\,\right] \nonumber\\
    &= \Gamma_m - \frac{\kappa}{8}\!\left[\,h_m + \Gamma_m h_m \Gamma_m\,\right], \label{eq:GD_fermionic_eqn_of_motion}
\end{align}
where, in the last step, we used the antisymmetry of $h_m$ and the identities $\Gamma_m^{-1} = \Gamma_m^T = -\Gamma_m$.
\subsubsection{Bosons}
\begin{figure}[t]
    \centering
    \includegraphics[width=\linewidth]{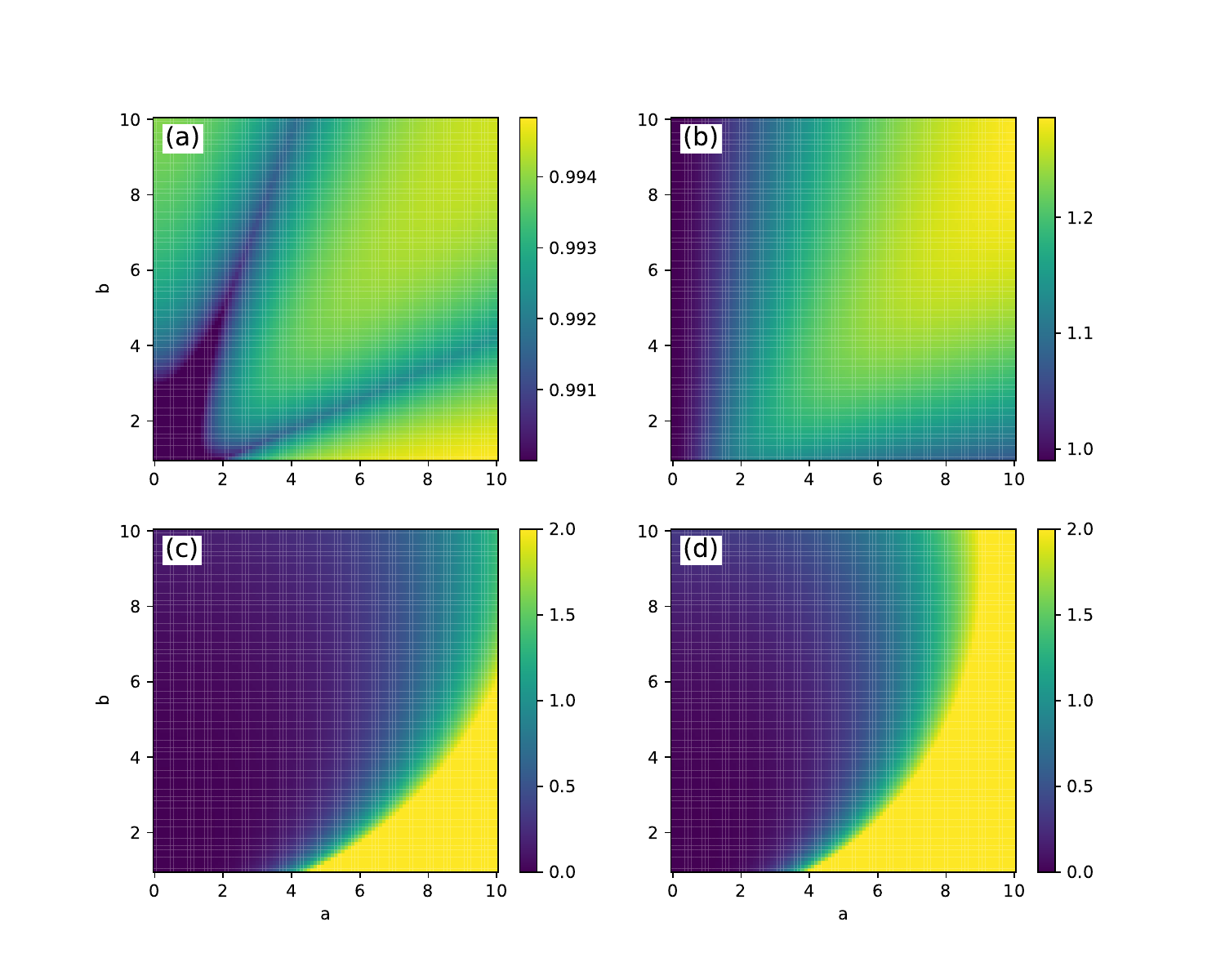}
    \caption{The variation of the length of the complete path of convergence for $\Delta_{R}$ and $\Gamma_b$, with different initial boson covariance matrices $\Gamma_b(a,b)$ for (a-b) GD and (c-d)ITE, respectively.}
    \label{fig:placeholder}
\end{figure}

Analogous to the fermions, we derive the GD equations for bosons. However, unlike the fermions, the bosons have two degrees of freedom in their variational parameters, namely the displacement vector $\Delta_R\in \mathds{R}^{2N_b}$, and the bosonic covariance matrix $\Gamma_b$, which is a real, symmetric, and symplectic matrix.\\
Similarly to the fermions, the gradient $\pdv{E}{\Gamma_b}$ needs to be projected on the appropriate tangent plane. To this end, we define the curve
\begin{align}
    \phi(t) = \Gamma_b e^{tA}, \qquad A \in \mathfrak{sp}(n,\mathds{R}),
\end{align}
where $\mathfrak{sp}(n,\mathds{R}) = \{A \in M(n,\mathds{R}) \, | \, \sigma A + A^{T}\sigma = 0 \}$ 
is the Lie algebra associated with the symplectic group $Sp(n,\mathds{R})$ . 
The corresponding tangent vector is 
$M = \left.\frac{d\phi(t)}{dt}\right|_{t=0} = \Gamma_b A$, 
which satisfies the constraint
\begin{align}
    M = \sigma (\Gamma_b^{T})^{-1} M^{T} (\Gamma_b^{T})^{-1} \sigma.
\end{align}

A valid projection of $h_b$ onto the tangent space is then given by
\begin{align}
    M' = \frac{\alpha}{2}\left[h_b + \Gamma_b \sigma h_b^{T} \sigma \Gamma_b \right], \qquad \alpha \in \mathds{R},
\end{align}
where $\alpha$ is chosen such that the update direction satisfies $\Tr(h_b^{T} M') > 0$, leading to
\begin{align}
    \alpha = \text{sgn}\!\left(\tfrac{1}{2}\Tr[h_b (h_b + \Gamma_b \sigma h_b \sigma \Gamma_b)]\right).
\end{align}
The resulting update equations for gradient descent are therefore
\begin{align}
    \Gamma_b &\rightarrow \Gamma_b - \frac{\alpha \kappa}{8}
      \left[h_b + \Gamma_b \sigma h_b \sigma \Gamma_b\right],\label{eq:GD_bosonic_covariance} \\
    \Delta_R &\rightarrow \Delta_R - \frac{\kappa}{2} h_{\Delta} \label{eq:GD_bosonic_displacement},
\end{align}
where $\kappa$ denotes the learning-rate or update parameter.
\subsubsection{Comparing to Imaginary-Time Evolution}

\begin{figure}[t]
    \centering
    \includegraphics[width=\linewidth]{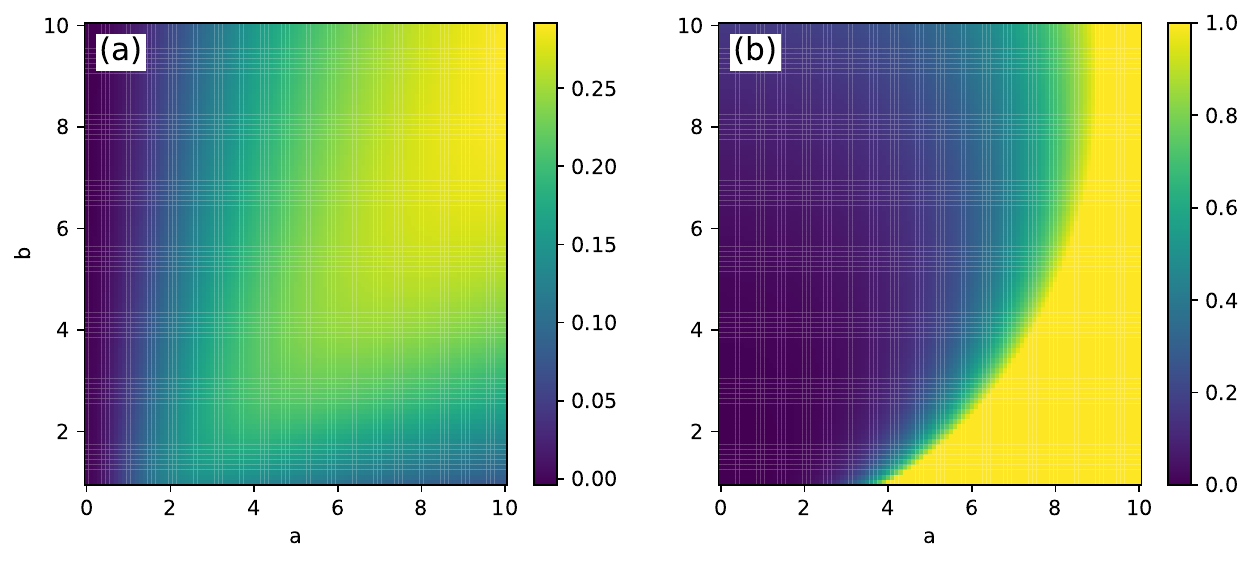}
    \caption{Difference of the length of convergence path for (a)$\Delta_R$ and (b)$\Gamma_b$ between ITE and GD for various initial boson covariance matrices $\Gamma_b(a,b)$ .}
    \label{fig:diff_lengths}
\end{figure}
Comparing the equations \ref{eq:ITE_fermionic_covariance} and \ref{eq:GD_fermionic_eqn_of_motion}, it is evident that GD and imaginary-time evolution (ITE) yield identical dynamics for fermionic systems. For bosonic systems, we can quantitatively compare the two methods, by restricting our attention to quadratic bosonic Hamiltonians,
\begin{align}
    H = \sum_{i,j} b_i^{\dagger} \, \omega_{ij} \, b_j,
\end{align}
where $b_i$ ($b_i^{\dagger}$) are the annihilation (creation) operators and $\omega$ is a Hermitian matrix. 
Since $\omega$ can be diagonalized by a linear transformation, and because the gradients $h_b$ and $h_{\Delta_R}$ transform covariantly under such transformations, it is convenient to perform the analysis in the diagonal basis of $\omega$. 
In this representation, the normal modes decouple, and hence the single-mode results directly generalize to multimode systems.

We therefore consider the single-mode Hamiltonian
\begin{align}
    H = \omega \, b^{\dagger} b,
\end{align}
and compare GD and ITE by tracking the path lengths of $\Delta_R$ and $\Gamma_b$ during evolution. 
The bosonic covariance matrix is parameterized as
\begin{align}
    \Gamma_b =
    \begin{bmatrix}
        a & b\\
        b & \frac{1+b^2}{a}
    \end{bmatrix}.
\end{align}
We examine the convergence behavior for different initial conditions for $\Gamma_b(a,b)$ at fixed $\Delta_R$. 
Figures \ref{fig:placeholder} and \ref{fig:diff_lengths} show the lengths of convergence paths for GD and ITE, as well as their difference. The length for the two ae defined by the respective inner products, euclidean for $\Delta_R$ and the trace inner product for $\Gamma_b$.\\
It is clear from Fig. \ref{fig:diff_lengths} that the convergence path in ITE is always longer than, or equal to, that in GD. 
Assuming comparable step sizes, this directly implies that ITE converges more slowly than GD. 
Physically, this slower convergence arises because when $\Gamma_b \neq \mathds{1}$, the ITE update direction deviates from the path of steepest descent in parameter space for $\Delta_R$, causing the dynamics for $\Delta_R$ to follow a longer trajectory toward the energy minimum.
\section{Conclusion}
In this paper, we re-investigated the relationship between imaginary-time evolution (ITE) and gradient descent (GD) and provided a detailed comparison between the two. Our analysis demonstrates that, contrary to the common assumption of equivalence, GD exhibits a faster convergence rate than ITE for bosonic systems. This insight suggests that advanced optimization techniques developed within the machine-learning community can be directly employed as efficient alternatives to ITE in variational approaches, without compromising convergence reliability. This result indicates how techniques from machine learning community can be used to accelerate computational many body physics. \\
Beyond classical simulations, the findings also raise intriguing questions for quantum algorithms, where quantum hybrid analogs of ITE and GD have been compared \cite{variational_ITE_quantum_computer}, and ITE was found to converge more rapidly. 
The apparent contrast between classical and quantum behavior invites further investigation into the underlying mechanisms and motivates the search for faster quantum implementations of GD-based optimization schemes.
\section{Acknowledgements}
The author acknowledges Prof.\ Dr.\ Falko Pientka for his steadfast support. The author also acknowledges Dr. Miguel Barbosa for interesting discussions that led to the development and actualization of this idea, and Dr. Jonas Profe for insightful discussions regarding the comparison of GD and ITE for bosonic Hamiltonians. Finally, the author expresses gratitude to Prof. Nitin Palan for his staunch support and valuable feedback during the preparation of this manuscript.
\bibliographystyle{unsrt}
\bibliography{references}

@Article{SciPostPhys:scalable_ITE_with_NN_quantum_states,
	title={{Scalable imaginary time evolution with neural network quantum states}},
	author={Eimantas Ledinauskas and Egidijus Anisimovas},
	journal={SciPost Phys.},
	volume={15},
	pages={229},
	year={2023},
	publisher={SciPost},
	doi={10.21468/SciPostPhys.15.6.229},
	url={https://scipost.org/10.21468/SciPostPhys.15.6.229},
}

@article{Shi2018:Variational_study_of_fermionic_and_bosonic_systems,
   author = {Tao Shi and Eugene Demler and J. Ignacio Cirac},
   doi = {10.1016/j.aop.2017.11.014},
   issn = {1096035X},
   journal = {Annals of Physics},
   title = {Variational study of fermionic and bosonic systems with non-Gaussian states: Theory and applications},
   volume = {390},
   year = {2018}
}

@article{Liang2018,
   author = {Xiao Liang and Huan Zhang and Sheng Liu and Yan Li and Yong Sheng Zhang},
   doi = {10.1038/s41598-018-34725-9},
   issn = {20452322},
   issue = {1},
   journal = {Scientific Reports},
   title = {Generation of Bose-Einstein Condensates’ Ground State Through Machine Learning},
   volume = {8},
   year = {2018}
}

@article{Wang:Hubbard_Holstein_model,
  title = {Zero-temperature phases of the two-dimensional Hubbard-Holstein model: A non-Gaussian exact diagonalization study},
  author = {Wang, Yao and Esterlis, Ilya and Shi, Tao and Cirac, J. Ignacio and Demler, Eugene},
  journal = {Phys. Rev. Res.},
  volume = {2},
  issue = {4},
  pages = {043258},
  numpages = {19},
  year = {2020},
  month = {Nov},
  publisher = {American Physical Society},
  doi = {10.1103/PhysRevResearch.2.043258},
  url = {https://link.aps.org/doi/10.1103/PhysRevResearch.2.043258}
}

@misc{Demler:Kondo_model,
      title={Kondo impurity in an attractive Fermi-Hubbard bath: Equilibrium and dynamics}, 
      author={Zhi-Yuan Wei and Tao Shi and J. Ignacio Cirac and Eugene A. Demler},
      year={2025},
      eprint={2501.05562},
      archivePrefix={arXiv},
      primaryClass={cond-mat.str-el},
      url={https://arxiv.org/abs/2501.05562}, 
}

@article{Costa:QMC_phonon_dispersion,
  title = {Phonon Dispersion and the Competition between Pairing and Charge Order},
  author = {Costa, N. C. and Blommel, T. and Chiu, W.-T. and Batrouni, G. and Scalettar, R. T.},
  journal = {Phys. Rev. Lett.},
  volume = {120},
  issue = {18},
  pages = {187003},
  numpages = {6},
  year = {2018},
  month = {May},
  publisher = {American Physical Society},
  doi = {10.1103/PhysRevLett.120.187003},
  url = {https://link.aps.org/doi/10.1103/PhysRevLett.120.187003}
}

@article{QMC_paper_2,
  title = {Two-dimensional $t\ensuremath{-}{t}^{\ensuremath{'}}$ Holstein model},
  author = {Ara\'ujo, Maykon V. and de Lima, Jos\'e P. and Sorella, Sandro and Costa, Natanael C.},
  journal = {Phys. Rev. B},
  volume = {105},
  issue = {16},
  pages = {165103},
  numpages = {8},
  year = {2022},
  month = {Apr},
  publisher = {American Physical Society},
  doi = {10.1103/PhysRevB.105.165103},
  url = {https://link.aps.org/doi/10.1103/PhysRevB.105.165103}
}

@article{QMC_paper_3,
  title = {Spectral Functions of the Holstein Polaron: Exact and Approximate Solutions},
  author = {Mitri\ifmmode \acute{c}\else \'{c}\fi{}, Petar and Jankovi\ifmmode \acute{c}\else \'{c}\fi{}, Veljko and Vukmirovi\ifmmode \acute{c}\else \'{c}\fi{}, Nenad and Tanaskovi\ifmmode \acute{c}\else \'{c}\fi{}, Darko},
  journal = {Phys. Rev. Lett.},
  volume = {129},
  issue = {9},
  pages = {096401},
  numpages = {6},
  year = {2022},
  month = {Aug},
  publisher = {American Physical Society},
  doi = {10.1103/PhysRevLett.129.096401},
  url = {https://link.aps.org/doi/10.1103/PhysRevLett.129.096401}
}

@article{DMFT_paper_1,
  title = {Electronic structure calculations with dynamical mean-field theory},
  author = {Kotliar, G. and Savrasov, S. Y. and Haule, K. and Oudovenko, V. S. and Parcollet, O. and Marianetti, C. A.},
  journal = {Rev. Mod. Phys.},
  volume = {78},
  issue = {3},
  pages = {865--951},
  numpages = {0},
  year = {2006},
  month = {Aug},
  publisher = {American Physical Society},
  doi = {10.1103/RevModPhys.78.865},
  url = {https://link.aps.org/doi/10.1103/RevModPhys.78.865}
}

@article{DMFT_paper_2,
  title = {Dynamical mean-field theory of strongly correlated fermion systems and the limit of infinite dimensions},
  author = {Georges, Antoine and Kotliar, Gabriel and Krauth, Werner and Rozenberg, Marcelo J.},
  journal = {Rev. Mod. Phys.},
  volume = {68},
  issue = {1},
  pages = {13--125},
  numpages = {0},
  year = {1996},
  month = {Jan},
  publisher = {American Physical Society},
  doi = {10.1103/RevModPhys.68.13},
  url = {https://link.aps.org/doi/10.1103/RevModPhys.68.13}
}

@article{DMRG_paper_1,
  title = {Density-matrix algorithms for quantum renormalization groups},
  author = {White, Steven R.},
  journal = {Phys. Rev. B},
  volume = {48},
  issue = {14},
  pages = {10345--10356},
  numpages = {0},
  year = {1993},
  month = {Oct},
  publisher = {American Physical Society},
  doi = {10.1103/PhysRevB.48.10345},
  url = {https://link.aps.org/doi/10.1103/PhysRevB.48.10345}
}

@article{DMRG_paper_2,
  title = {Density-matrix renormalization-group study of the polaron problem in the Holstein model},
  author = {Jeckelmann, Eric and White, Steven R.},
  journal = {Phys. Rev. B},
  volume = {57},
  issue = {11},
  pages = {6376--6385},
  numpages = {0},
  year = {1998},
  month = {Mar},
  publisher = {American Physical Society},
  doi = {10.1103/PhysRevB.57.6376},
  url = {https://link.aps.org/doi/10.1103/PhysRevB.57.6376}
}

@article{DMRG_paper_3,
  title = {Density Matrix Approach to Local Hilbert Space Reduction},
  author = {Zhang, Chunli and Jeckelmann, Eric and White, Steven R.},
  journal = {Phys. Rev. Lett.},
  volume = {80},
  issue = {12},
  pages = {2661--2664},
  numpages = {0},
  year = {1998},
  month = {Mar},
  publisher = {American Physical Society},
  doi = {10.1103/PhysRevLett.80.2661},
  url = {https://link.aps.org/doi/10.1103/PhysRevLett.80.2661}
}

@article{DFT_paper_1,
doi = {10.1088/0953-8984/27/39/393001},
url = {https://doi.org/10.1088/0953-8984/27/39/393001},
year = {2015},
month = {sep},
publisher = {IOP Publishing},
volume = {27},
number = {39},
pages = {393001},
author = {Carrascal, D J and Ferrer, J and Smith, J C and Burke, K},
title = {The Hubbard dimer: a density functional case study of a many-body problem},
journal = {Journal of Physics: Condensed Matter}
}

@article{variational_ITE_quantum_computer,
   author = {Sam McArdle and Tyson Jones and Suguru Endo and Ying Li and Simon C. Benjamin and Xiao Yuan},
   doi = {10.1038/s41534-019-0187-2},
   issn = {20566387},
   issue = {1},
   journal = {npj Quantum Information},
   title = {Variational ansatz-based quantum simulation of imaginary time evolution},
   volume = {5},
   year = {2019}
}
\end{document}